\documentclass[twocolumn,aps,prb,floats,floatfix,superscriptaddress]{revtex4-2}

\usepackage{amsmath}
\usepackage{graphicx}
\usepackage{dcolumn}
\usepackage[english]{babel}

\usepackage{color}
\usepackage{float}

\begin{document}

\title{Sensitivity of x-ray absorption at $5d$ edges of high-valent light actinides to crystal-field strength and covalency effects}

\author{J. Vegelius}
\altaffiliation[Present ]{address: Department of Medical Sciences, Uppsala University, Sweden}
\affiliation{Condensed Matter Physics of Energy Materials, X-ray Photon Science, Department of Physics and Astronomy, Uppsala University, P.O. Box 516, SE-751 20 Uppsala, Sweden}
\author{D. K. Shuh}
\affiliation{Chemical Sciences Division, Lawrence Berkeley National Laboratory, MS 70A3307, One Cyclotron Road, Berkeley, CA 94720, USA}
\author{S. M. Butorin}
\altaffiliation{Corresponding author}
\affiliation{Condensed Matter Physics of Energy Materials, X-ray Photon Science, Department of Physics and Astronomy, Uppsala University, P.O. Box 516, SE-751 20 Uppsala, Sweden}



\begin{abstract}
The $5f$ states were probed in ThO$_2$ and U(VI) and U(V) oxides using x-ray absorption spectroscopy at the actinide $O_{4,5}$ edges. Measured data were analyzed by several approaches including atomic, crystal-field theory and the Anderson impurity model to take into account the hybridization of actinide valence states with oxygen $2p$ states. For Th(IV), the $5f$ states are mainly affected by the crystal-field interaction with the closest neighbors as can be seen from the corresponding spectrum of ThO$_2$. In turn, for U(VI) and U(V) oxides, the U $5f$-O $2p$ hybridization and  high degree of covalency in the chemical bonding play a decisive role increasing the $5f$ occupancy and consequently  governing the ground and excited state properties. That is additionally illustrated by results from the calculations of the U $5d$ x-ray photoemission spectra for U(V) in the lattice of La-doped UO$_2$.
\end{abstract}


\maketitle

\section{Introduction}
The role of the local atomic environment in defining the properties of high-valent actinide compounds is enhanced due to a larger degree of the $5f$ delocalization and stronger interactions induced by close neighbors/ligands of the actinide atom, such as a crystal-field potential. The direct probing of the $5f$ states helps to characterize the electronic structure and properties of the actinide compounds more efficiently, which can be done by spectroscopic methods with strict selection rules, such as, for example, x-ray absorption spectroscopy (XAS). For actinides, the conventional XAS measurements are usually performed at $3d$ ($M_{4,5}$), $4d$ ($N_{4,5}$) and $5d$ ($O_{4,5}$) edges to study transitions to the $5f$ states. However, a large core-hole lifetime broadening \cite{Campbell} of the actinide $M_{4,5}$ and $N_{4,5}$ XAS spectra does not allow one to directly resolve any structures besides the main lines. The core-hole lifetime broadening can be reduced by employing the HERFD-XAS (high energy resolution fluorescence detection x-ray absorption spectroscopy) method, as demonstrated by experiments at actinide $M_{4,5}$ edges (see e.g., review \cite{Kvashnina_review}), but this type of measurements requires access to crucial additional equipment, such as a x-ray emission spectrometer.

At the $O_{4,5}$ thresholds, the main absorption edge is very broad due to so-called autoionization processes, however for light actinides, the "pre-edge" structures are observed at several eVs below the main edge \cite{Aono,Iwan,Adachi,Fujimori,Kalkowski,Cox,Butorin_JES,Nelson,Butorin_handbook,Moore,Butorin_nano,Butorin_AnalChem}, which are pushed towards low energies by a strong $5d$-$5f$ interaction. Such structures become almost immersed in the main edge for Pu \cite{Tobin,Moore,Butorin_nano} and are not observed for actinides with higher atomic numbers, such as Am \cite{Moore,Wiss,Butorin_Am_ox,Butorin_Am2O3}, Cm \cite{Moore,Kvashnina_Cm2O3} and Cf \cite{Muller}. Here, we show that "pre-edge" structures can be sensitive to the crystal-field strength and hybridization effects in high-valent light actinide compounds and therefore can be used for the characterization of the chemical bonding.

The XAS measurements at the shallow $5d$ edges of actinides need to be performed in ultra-high vacuum which makes it difficult to carry out such experiments at synchrotron radiation facilities due to safety restrictions. In this case, the use of high-resolution electron-energy-loss spectroscopy (EELS) at laboratories specialized in actinide research would be of advantage, because samples with much higher activities can be studied there. That is also facilitated by rather high EELS cross section at $O_{4,5}$ edges \cite{Nelson,Moore,Wiss,Cukier,Moser,Keller,Rice,Bradley,Degueldre} compared to other edges. However, in order to be successful, a good understanding of the factors affecting $O_{4,5}$ XAS is necessary. Therefore, in the present paper, we analyze the $O_{4,5}$ XAS spectra of ThO$_2$ and U(VI), U(V) oxides using atomic, crystal-field multiplet theory and the Anderson impurity model (AIM) \cite{Anderson}.

\section{Experimental}
The ThO$_2$ sample (99.99\% purity) was acquired from Alfa Aesar. The U$_3$O$_7$ samples was prepared as described in Ref. \cite{Allen}. Flakes of ThO$_2$ and U$_3$O$_7$ were glued on the sample holders using a silver epoxy to provide a good contact for the total electron yield (TEY) measurements in the drain current mode. The Th and U $O_{4,5}$ ($5d\rightarrow5f,7p$ transitions) XAS data of ThO$_2$ were recorded at beamline 5.3.1 of the MAXlab \cite{Denecke} equipped with the SX-700 monochromator. The monochromator resolution was set to $\sim$50 meV at 100 eV (115 eV) during measurements on ThO$_2$ (U$_3$O$_7$), respectively.

The UO$_2$ and UO$_3$ samples were pellets made from the 99.8\%-purity powder bought from Alfa Aesar. They were attached to a specially designed sample holder, which is described in Refs. \cite{Butorin_nano,Smiles}, to ensure that no contamination will be left in the experimental chamber after measurements. Experiments in the energy range of the U $O_{4,5}$ x-ray absorption thresholds of uranium oxides were performed at beamline 7.0.1 of the Advanced Light Source, Lawrence Berkeley National Laboratory employing a spherical grating monochromator \cite{Warwick}. The U 5$d$ XAS data were measured in the TEY mode using the drain current on the sample and the channeltron. The monochromator resolution was set to $\sim$50 meV at 115 eV during measurements at the U 5$d$ edges.

\section{Computational details}
The XAS spectra at the actinide $5d$ edges were calculated by taking into account the inhomogeneous broadening of the spectral transitions (due to differences in the core-hole lifetime of various core-excited states) and Fano effect. The $5d$ core-hole lifetimes were estimated using the formalism described in Ref. \cite{Ogasawara}. In our calculations, in addition to the interactions for the free ion, as used in Ref. \cite{Ogasawara}, the crystal-field splittings in the $5f$ shell and the hybridization effects between actinide and oxygen states were taken into account using the AIM approach.

The typical Hamiltonian in AIM, which includes the $5f$ and core $5d$ states on a single actinide ion and the O $2p$ states, can be written as
\begin{eqnarray}
H&=&\varepsilon_{5f}\sum_{\gamma} a^{\dag}_{5f}(\gamma)a_{5f}(\gamma) \nonumber  \\
       &+&
\varepsilon_{5d}\sum_{\mu} a^{\dag}_{5d}(\mu)a_{5d}(\mu) \nonumber  \\
       &+&
\sum_{\sigma,\gamma} \varepsilon_{\upsilon}(\sigma)a^{\dag}_{\upsilon}(\sigma,\gamma)a_{\upsilon}(\sigma,\gamma) \nonumber  \\
       &+& U_{ff}\sum_{\gamma>\gamma^{\prime}}a^{\dag}_{5f}(\gamma)a_{5f}(\gamma)a^{\dag}_{5f}(\gamma^{\prime})a_{5f}(\gamma^{\prime})  \nonumber\\
       &-&
       U_{fc}\sum_{\gamma,\mu}a^{\dag}_{5f}(\gamma)a_{5f}(\gamma)a^{\dag}_{5d}(\mu)a_{5d}(\mu) \nonumber\\
       &+&
       \frac{V}{\sqrt{N}}\sum_{\sigma,\gamma} [(a^{\dag}_{\upsilon}(\sigma,\gamma)a_{5f}(\gamma) + a^{\dag}_{5f}(\gamma)a_{\upsilon}(\sigma,\gamma)] \nonumber\\
       &+&
       H_{multiplet},
\end{eqnarray}
where $\varepsilon_{5f}$, $\varepsilon_{5d}$, and $\varepsilon_{\upsilon}$ are one-electron energies of the $5f$, core $5d$ and valence band levels, respectively, and $a^{\dag}_{5f}(\gamma)$, $a^{\dag}_{5d}(\mu)$, $a^{\dag}_{\upsilon}(\sigma,\gamma)$ are electron creation operators at these levels with combined indexes $\gamma$ and $\mu$ to represent the spin and orbital states of the $5f$, $5d$ and valence-band electrons, $\sigma$ is the index of the $N$ discrete energy levels in the O $2p$ band (bath states). $U_{fc}$ is the core $5d$ hole potential acting on the $5f$ electrons. $V$ is the hybridization strength (or hopping term) between the $5f$ states and states of the O $2p$ band. The $\varepsilon_{\upsilon}(\sigma)$ is represented by the $N$ discrete levels/bath states in the form
\begin{eqnarray}
\varepsilon_{\upsilon}(\sigma)=\varepsilon_{\upsilon}^0-\frac{W}{2}+\frac{W}{N}(\sigma-\frac{1}{2}),~\sigma=1,...,N,
\end{eqnarray}
where $\varepsilon_{\upsilon}^0$ and $W$ are the center and width of the O $2p$ band, respectively.

$H_{multiplet}$ includes the electrostatic and spin-orbit interactions on the actinide ion \cite{Butorin_UO2,Butorin_3d4fRIXS} as well as interactions leading to the crystal-field splittings of the $5f$ shell. The interactions between $5f$ electrons and between a core $5d$ hole and $5f$ electrons are described in terms of Slater integrals $F^{2,4,6}(5f,5f)$, $F^{2,4}(5d,5f)$ and $G^{1,3,5}(5d,5f)$, while the spin-orbit interactions for the $5f$ and core $5d$ states are described with coupling constants $\zeta(5f)$ and $\zeta(5d)$, respectively. Actinide $5f$-O $2p$ charge-transfer energy $\Delta_{5f}$ is taken as $\Delta_{5f}=\varepsilon_{5f}-\varepsilon_{\upsilon}^0$.

However, ThO$_2$ is a special case. One-electron energy $\varepsilon_{6d}$ is lower than $\varepsilon_{5f}$, the bottom of the conduction band is dominated by Th $6d$ states so that the Th $6d-$O $2p$ charge transfer is expected to be more energetically favorable. That needs to be taken into account for the Th $5d$ XAS process due to Th $6d$-Th $5f$-O $2p$ hybridization as it was done in calculations of Th $3d$ XAS of ThO$_2$ (Ref. \cite{Butorin_PNAS}). Additional parameters were introduced, such as Th $6d-$O $2p$ charge-transfer energy $\Delta_{6d}$($=\varepsilon_{6d}-\varepsilon_{\upsilon}^0$, Coulomb interactions between $5f$ electrons and $6d$ electrons $U_{fd}$ and between a core-hole and $6d$ electrons $U_{dc}$.

The \textit{ab-initio} values of Slater integrals, spin-orbit coupling constants and matrix elements were obtained with the TT-MULTIPLETS package which combines Cowan's atomic multiplet program \cite{Cowan} (based on the Hartree-Fock method with relativistic corrections) and Butler's point-group program \cite{Butler}, which were modified by Thole \cite{Thole}, as well as the charge-transfer program written by Thole and Ogasawara \cite{Kotani_4dXAS_CeO2}.

To compare with the experimental data, it is usually necessary to uniformly shift the calculated spectra on the energy scale because it is difficult to accurately reproduce the absolute energies in this type of calculations.

\section{Results and discussion}
\begin{figure}[t]
\includegraphics[width=\columnwidth]{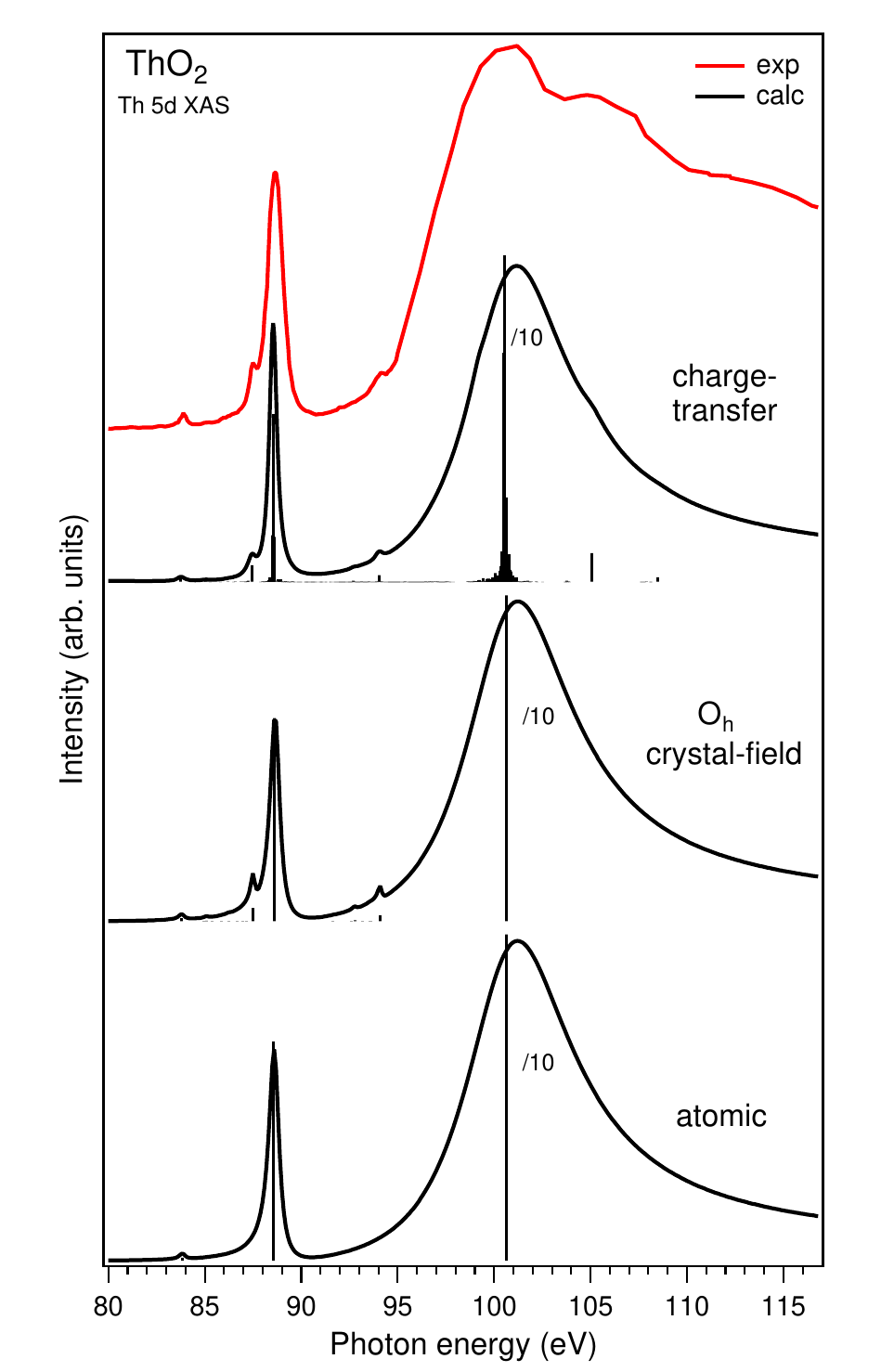}.
\caption{Experimental and calculated XAS spectra at Th $O_{4,5}$ edges of ThO$_2$. The spectra are calculated using the atomic and crystal-field multiplet theory for the Th(IV) ion and Anderson impurity model, respectively. Note, that the most intense multiplet pole in each calculated spectrum was reduced by 10 times for a better presentation. \label{5dXAS_ThO2}}
\end{figure}

The experimental Th $O_{4,5}$ XAS spectrum of ThO$_2$ is displayed in Fig.~\ref{5dXAS_ThO2}. For ThO$_2$, the autoinization processes can be characterized by two decay channels following the $5d^{10}5f^{0}\rightarrow5d^{9}5f^{1}$ excitation. The $5d\rightarrow5f$ absorption and $5d\rightarrow\varepsilon{f}$ ionization are competing routes coupled to each other by the $\langle5d^{9}5f^{1}|1/r|5d^{9}5f^{0}\varepsilon{f}\rangle$ configuration interaction (CI). The decay of the excited $5d^{9}5f^{1}$ states via $\langle5d^{9}5f^{1}|1/r|5d^{9}5f^{0}\varepsilon{f}\rangle$ can be called as the $5f\rightarrow\varepsilon{f}$ tunneling channel \cite{Ogasawara}. The $5d^{9}5f^{1}$ core excitations can also decay via the $5d-5f6(s,p)$ Coster-Kronig channel so that the excited states are coupled to the $6(s,p)^{-1}$ ionization continua by the $\langle5d^{9}5f^{1}|1/r|5d^{10}5f^{0}6(s, p)^{-1}\varepsilon{l}\rangle$ CI process. The broadening of the $5f$ edge which is highly inhomogeneous can be evaluated by calculating these CI matrix elements. For $5f^1$-systems, the $5d-5f5f$ super-Coster-Kronig decay channel becomes open via $\langle5d^{9}5f^{2}|1/r|5d^{10}5f^{0}\varepsilon{l}\rangle$. However, for light lanthanides and actinides, the tunneling decay channel is the dominating process \cite{Ogasawara}.

Fig.~\ref{5dXAS_ThO2} also shows the calculated Th $O_{4,5}$ XAS spectra of ThO$_2$. Besides using the atomic-multiplet theory as in Refs. \cite{Ogasawara_5dXAS,Gupta}, the crystal-field interaction and hybridization of Th states with oxygen states were included in the calculations for the Th(IV) system. Similar crystal-field and AIM parameter values were applied as in the calculations of the Th $M_{4}$ XAS spectrum of ThO$_2$ (Ref. \cite{Butorin_PNAS}), except for the values of the Slater integrals. It has been established (see e.g., \cite{Ogasawara_5dXAS}) that the Slater integrals describing the interactions of the $f$ electrons with core holes in shallow levels, such as $4d$ for lanthanides and $5d$ for actinides, need to be additionally scaled down from their \textit{ab-initio} Hartree-Fock values for a description of the XAS spectra at those levels. In our calculations, Slater integrals $F^{k}$ for the ground state and $F^{k}$, $G^{k}$ and $R^{k}$ for the excited state were reduced to 80\%, 75\%, 60\% and 70\%, respectively.

An inspection of Fig.~\ref{5dXAS_ThO2} reveals that the Th(IV) $O_{4,5}$ XAS spectrum calculated using atomic multiplet theory has contributions from only three transitions to the ${^3}P_1$ (at $\sim$83.9 eV), ${^3}D_1$ (at $\sim$88.6 eV) and ${^1}P_1$ (at $\sim$100.6 eV) states. An application of the crystal field of cubic symmetry with Wybourne's crystal-field parameters $B^{4}_{0}$=-1.3 eV and $B^{6}_{0}$=0.55 eV (which are the same as in the Th $3d-4f$ resonant inelastic x-ray scattering (RIXS) calculations \cite{Butorin_3d4fRIXS} of ThO$_2$) leads to an appearance of additional transitions as compared to the atomic multiplet case. These transitions become allowed due to so-called $J$-mixing (see e.g., \cite{Butorin_JES}). The most noticable results of the crystal-field interaction is a split of the 88.6-eV peak and a small new structure at 94.1 eV which are in good correspondence with the features of the experimental spectrum.

In AIM calculations for the Th(IV) system, the ground (core-excited) state for the XAS process was described as a combination of the $5f^{0}$,  $5f^{0}6d^{1}\underline{\upsilon}^{1}$ and $5f^{1}\underline{\upsilon}^{1}$ ($5d^{9}5f^{1}$,  $5d^{9}5f^{1}6d^{1}\underline{\upsilon}^{1}$ and $5d^{9}5f^{2}\underline{\upsilon}^{1}$) configurations. The model parameters had the following values: $\Delta_{6d}=5.0$, $\Delta_{5f}=8.0$, $U_{ff}=4.0$, $U_{fc}=5.0$, $U_{dc}-U_{df}=1.0$, $V=1.1$ in units of eV. In the core-excited state, $V$ was reduced to 64\% of its value (reduction factor $\kappa=0.64$) to take into account the configuration dependence \cite{Gunnarsson} of $V$. $U_{ff}\simeq{4.0}$ eV was deduced from the fit of the Th $4f$ x-ray photoemission data \cite{Kotani}. In addition, according to estimations in \cite{Kotani_AdvPhys}, $U_{df}$ can be as large as $U_{df}\simeq{U_{fc}/2}$.

In the limit of $V\rightarrow0$, the difference between the configuration averaged energies for the ground state was $E(5f^{0}6d^{1}\underline{\upsilon}^{1})-E(5f^{0})=\Delta_{6d}$ and $E(5f^{1}\underline{\upsilon}^{1})-E(5f^{0})=\Delta_{5f}$ and for the core-excited state it was $E(5d^{9}5f^{1}6d^{1}\underline{\upsilon}^{1})-E(5d^{9}5f^{1})=\Delta_{6d}+U_{df}-U_{dc}$ and $E(5d^{9}5f^{2}\underline{\upsilon}^{1})-E(5d^{9}5f^{1})=\Delta_{5f}+U_{ff}-U_{fc}$. In addition to the crystal-field interaction for the $5f$ shell, the crystal-field strength for the Th $6d$ shell was set to $10Dq=-4.0$ eV in the AIM calculations. As a result, the AIM-calculated spectrum (Fig.~\ref{5dXAS_ThO2}) indicates that the charge-transfer satellites contribute to the broad experimental structure at around 105.5 eV. This structure and other structures at higher photon energies also originate from the $5d\rightarrow7p$ transitions which are not included in our calculations. As in case of the Th $M_4$ XAS calculations \cite{Butorin_PNAS}, the Th $6d$ and $5f$ covalent contribution in the ground state of ThO$_2$ was estimated to be 0.20 and 0.11 electrons, respectively, thus refuting the ionic character of the chemical bonding in this dioxide. However, the $5f$ states are mainly affected by the crystal-field interaction with the closest neighbors because taking into account the hybridization of the Th states with O $2p$ states in the calculations does not lead to significant changes in the pre-edge fine spectral structure.

\begin{figure}[t]
\includegraphics[width=\columnwidth]{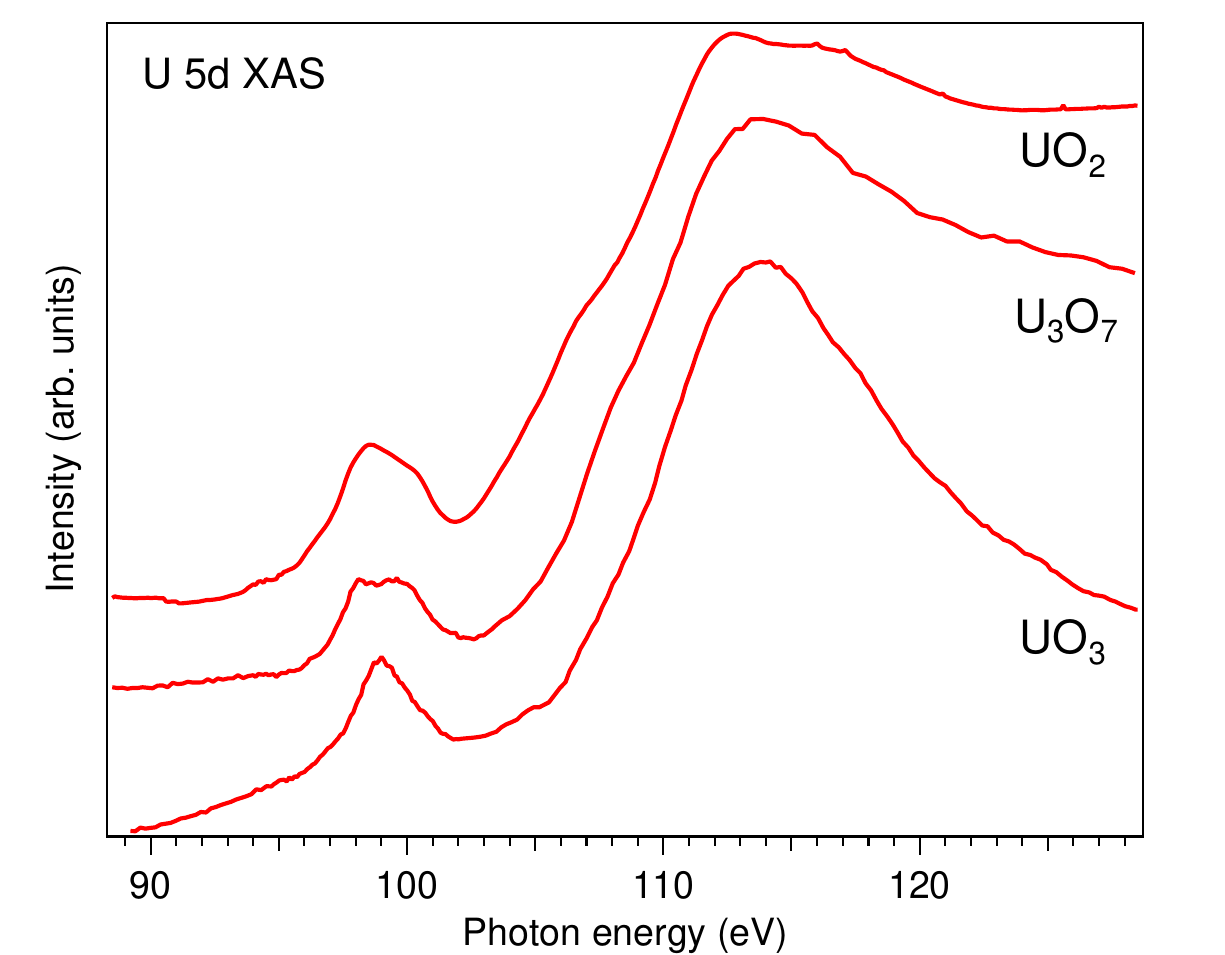}
\caption{U $O_{4,5}$ XAS spectra of uranium oxides recorded in the TEY mode. \label{U5dXAS_Uox}}
\end{figure}

Fig.~\ref{U5dXAS_Uox} displays the U $O_{4,5}$ XAS spectra of UO$_3$, U$_3$O$_7$ and UO$_2$. The spectra of UO$_3$ and UO$_2$ are similar to those published by Kalkowski \textit{et al.} \cite{Kalkowski}. The different energies of the main maximum of the "giant resonance" edge in these oxides can be related to the chemical shift due to different oxidation state of uranium: U(VI) in UO$_3$, U(IV)+2U(V) in U$_3$O$_7$ (see Ref. \cite{Leinders}) and U(IV) in UO$_2$. Furthermore, the differences in the shape of the pre-edge structures are clearly observed. In addition, the shoulder at $\sim$107 eV in the spectrum of UO$_2$ is barely seen in the spectrum of U$_3$O$_7$ and cannot be observed for UO$_3$.

To outline the $O_{4,5}$ XAS differences between ThO$_2$ and UO$_3$, where both actinides are in the formal $5f^0$ state, the AIM-calculations were also carried for the UO$_3$ case. To take into account an expected strong U $5f$-O $2p$ hybridization in the U(VI) system, the ground (core-excited) state was described by a mixture of $5f^{0}$, $5f^{1}\underline{\upsilon}^{1}$ and $5f^{2}\underline{\upsilon}^{2}$ ($5d^{9}5f^{1}$, $5d^{9}5f^{2}\underline{\upsilon}^{1}$ and $5d^{9}5f^{3}\underline{\upsilon}^{2}$) configurations. The reduction factors for the \textit{ab-initio} Hartree-Fock values of Slater integrals were the same as in case of the Th $O_{4,5}$ XAS calculations for ThO$_2$. The crystal field acting on the U(VI) ions in UO$_3$ was approximated to be of octahedral symmetry with $B^{4}_{0}$=1.46 eV and $B^{6}_{0}$=0.23 eV. The AIM-parameters were almost the same as those used in the $3d$-$5f$ (valence-to-core) RIXS calculations \cite{Nakazawa} for UO$_3$ to reproduce the experimental data \cite{Butorin_PRL,Zatsepin}: $\Delta_{5f}=0.5$, $U_{ff}=3.0$ and $V=1.5$ in units of eV. For the shallow $5d$ core-hole, $U_{fc}$ was taken to be 4.5 eV and reduction factor $\kappa$ for $V$ in the core-excited state was 0.7.

\begin{figure}[b]
\includegraphics[width=\columnwidth]{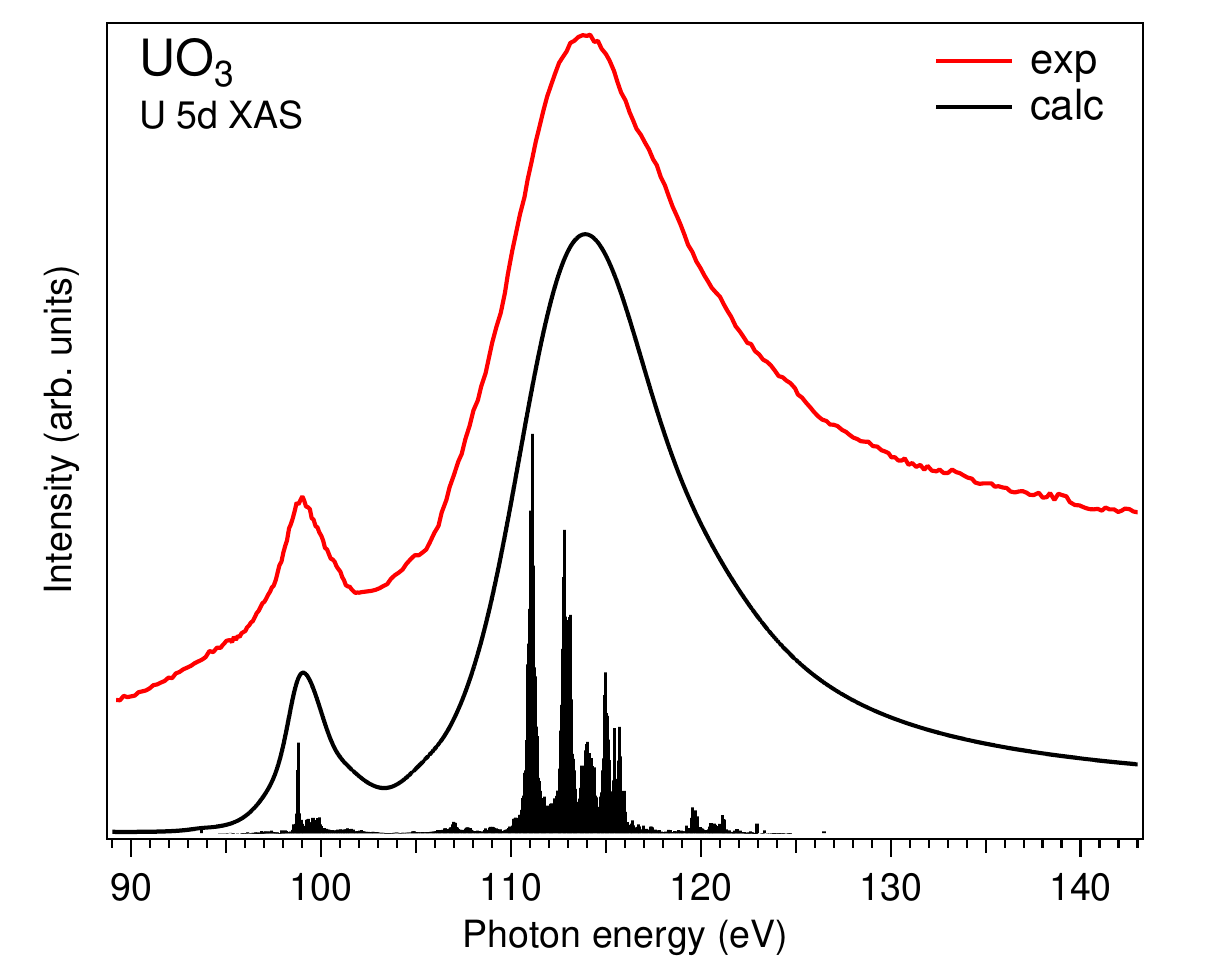}
\caption{Experimental and calculated XAS spectra at U $O_{4,5}$ edges of UO$_3$. \label{UO3_5d}}
\end{figure}

In the limit of $V\rightarrow0$, the differences between the configuration averaged energies can be written as $E(5f^{n+1}\underline{\upsilon}^{1})-E(5f^{n})=\Delta_{5f}$,  $E(5f^{n+2}\underline{\upsilon}^{2})-E(5f^{n+1}\underline{\upsilon}^{1}))=\Delta_{5f}+U_{ff}$ for the ground state and $E(5d^{9}5f^{n+2}\underline{\upsilon}^{1})-E(5d^{9}5f^{n+1})=\Delta_{5f}+U_{ff}-U_{fc}$,  $E(5d^{9}5f^{n+3}\underline{\upsilon}^{2})-E(5d^{9}5f^{n+2}\underline{\upsilon}^{1})=\Delta_{5f}+2U_{ff}-U_{fc}$ for the core-excited state, where $n=0$ for the U(VI) ion.

\begin{figure}[t]
\includegraphics[width=\columnwidth]{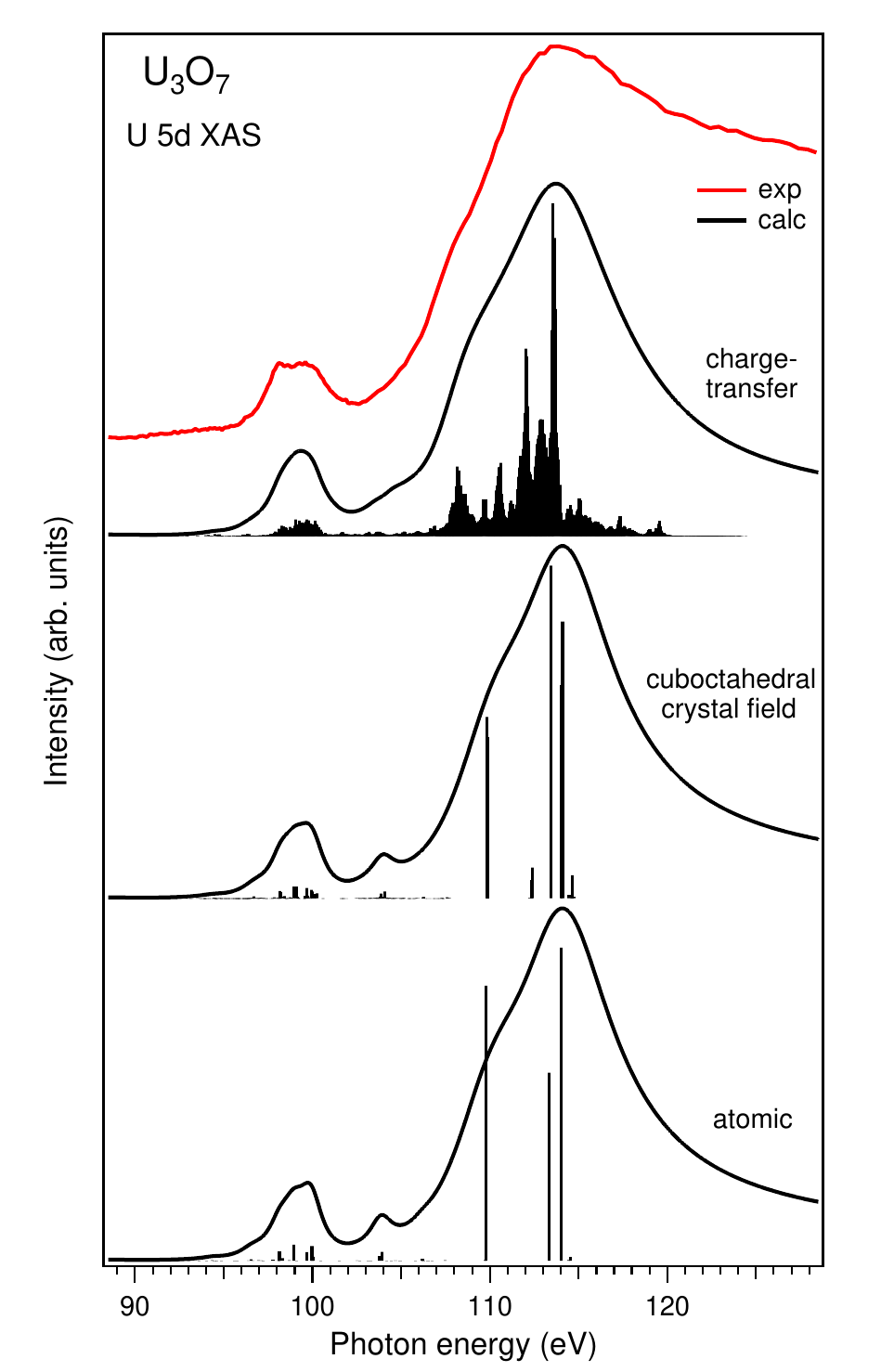}
\caption{Comparison of the U $O_{4,5}$ XAS spectrum of U$_3$O$_7$ with calculated results for the U(V) system. The $O_{4,5}$ XAS spectra are calculated using the atomic and (cuboctahedral) crystal-field multiplet theory and Anderson impurity model, respectively. \label{U3O7_5d}}
\end{figure}

Fig.~\ref{UO3_5d} shows a comparison of the experimental and calculated U $O_{4,5}$ XAS spectra of UO$_3$. The pre-edge structure at around 99.0 eV appears to be significantly broader than the corresponding one in the Th $O_{4,5}$ XAS spectrum of ThO$_2$. That is a result of the superposition of the transitions to the states of the $5d^{9}5f^{1}$ and $5d^{9}5f^{2}\underline{\upsilon}^{1}$ configurations which are strongly coupled via $V$. The region between $\sim$100 and $\sim$108 eV is contributed to by transitions to the states of the $5d^{9}5f^{3}\underline{\upsilon}^{2}$ configuration. As a whole, the $5f$ occupancy in UO$_3$ was estimated to be 0.97 electrons in this calculation, thus indicating the high degree of covalency for the chemical bonding.

The derived U(VI) $5f$ occupancy in UO$_3$ due to covalency is higher than that for the U(VI) ion residing in the UO$_2$ lattice as a result of the doping by Nd (see Refs. \cite{Bes,Kvashnina_review}) or for U(VI) in uranate Pb$_3$UO$_6$ (Ref. \cite{Butorin_uranates}). That is because U $5f-$O $2p$ charge-transfer energy $\Delta_{5f}$ was estimated to be 0.5 eV in UO$_3$ versus 2.0 eV in (U,Nd)O$_2$ and Pb$_3$UO$_6$.

\begin{figure}[t]
\includegraphics[width=\columnwidth]{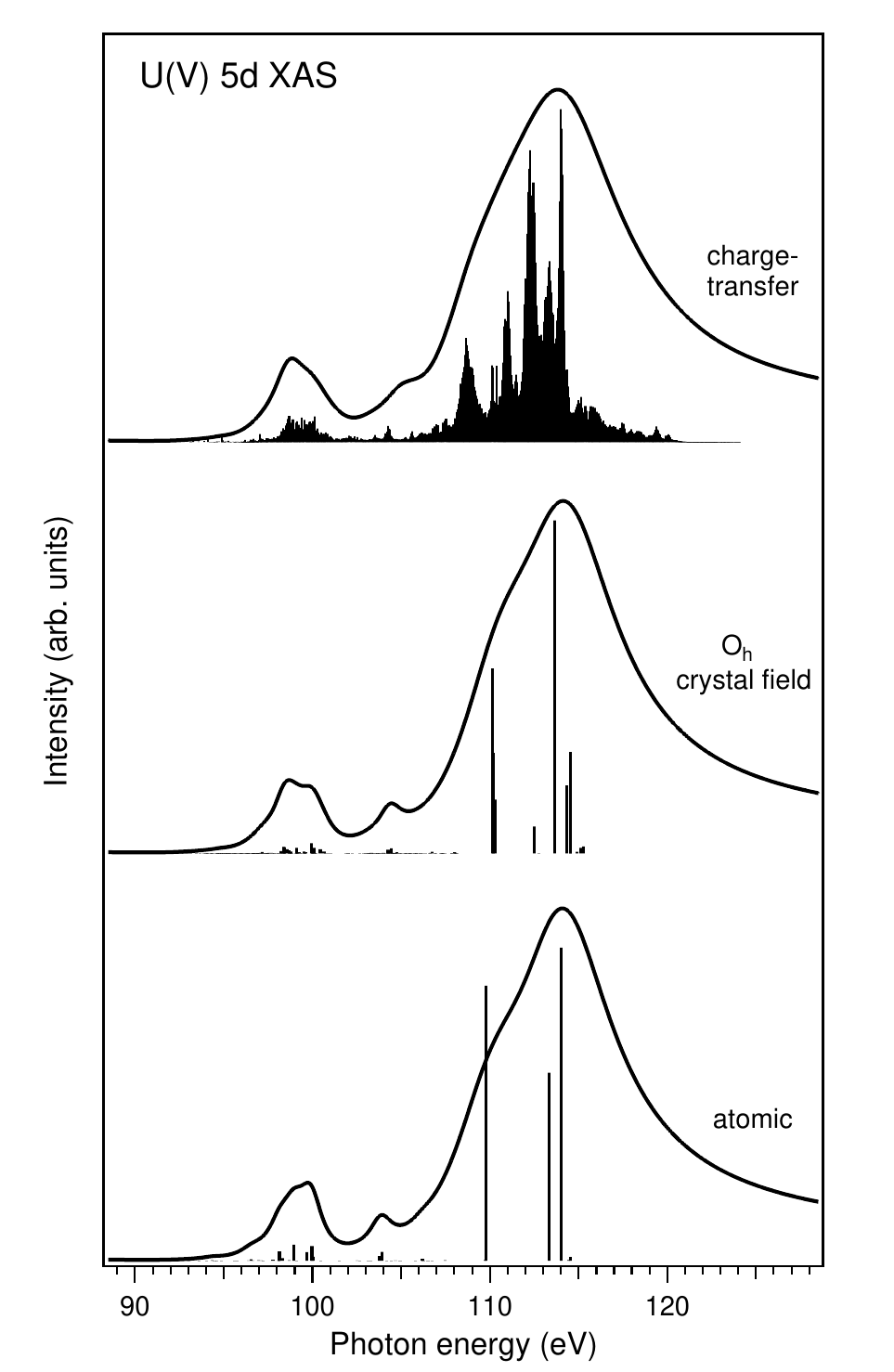}
\caption{Calculated $O_{4,5}$ XAS spectra using the atomic and (octahedral) crystal-field multiplet theory for the U(V) ion and Anderson impurity model, respectively. \label{U5_Oh_CT}}
\end{figure}

The same approaches were employed to calculate the $O_{4,5}$ XAS spectra for the U(V) system. The calculated spectra are shown in Fig.~\ref{U3O7_5d} and compared with the measured $O_{4,5}$ XAS spectrum of U$_3$O$_7$ since 2/3 of U sites are in the U(V) chemical state in this oxide \cite{Leinders}. In the calculations, the reduction factors for the \textit{ab-initio} Hartree-Fock values of the Slater integrals for the U(V) ion were the same as in the ThO$_2$ and UO$_3$ cases. Since U$_3$O$_7$ contains the cuboctahedral oxygen clusters \cite{Garrido,Bazarkina}, the cuboctahedral crystal field was applied in the calculations with Wybourne's parameters values $B^{4}_{0}$=-0.52 eV and $B^{6}_{0}$=-0.48 eV. The values are related to those of UO$_2$ (cubic crystal field) and estimated according to equations in Ref. \cite{Butorin_AnIII}. The AIM-parameters had the values similar to the case of U(V) in the lattice of the La-doped UO$_2$ (Ref. \cite{Butorin_La-UO2}): $\Delta_{5f}=3.5$, $U_{ff}=3.5$ and $V=1.2$ in units of eV. Reduction factor $\kappa$ for $V$ in the core-excited state was 0.8. For the shallow $5d$ core-hole, $U_{fc}$ was taken to be 5.0 eV. The ground (core-excited) state was described as a mixture of the $5f^{1}$, $5f^{2}\underline{\upsilon}^{1}$ and $5f^{3}\underline{\upsilon}^{2}$ ($5d^{9}5f^{2}$, $5d^{9}5f^{3}\underline{\upsilon}^{1}$ and $5d^{9}5f^{4}\underline{\upsilon}^{2}$) configurations.

The calculated U(V) $O_{4,5}$ XAS spectra using the atomic and crystal-field multiplet approached reveal the $\sim$104-eV structure which is not observed in the recoded U $O_{4,5}$ XAS spectrum of U$_3$O$_7$. On the other hand, this structure is almost completely smeared out in the AIM-calculated spectrum, thus making the spectral profile closer to that of the experimental spectrum. This indicates the importance of taking into account the U $5f$-O $2p$ hybridization and charge-transfer effects. Nevertheless, a small difference in the shape of the pre-edge structure around 98.1 eV is observed between the AIM-calculated and experimental spectrum (Fig.~\ref{U3O7_5d}). This difference can be attributed to the U(IV) contribution to the spectrum of U$_3$O$_7$ (see the spectrum of UO$_2$ in Fig.~\ref{U5dXAS_Uox}).

\begin{figure}[t]
\includegraphics[width=\columnwidth]{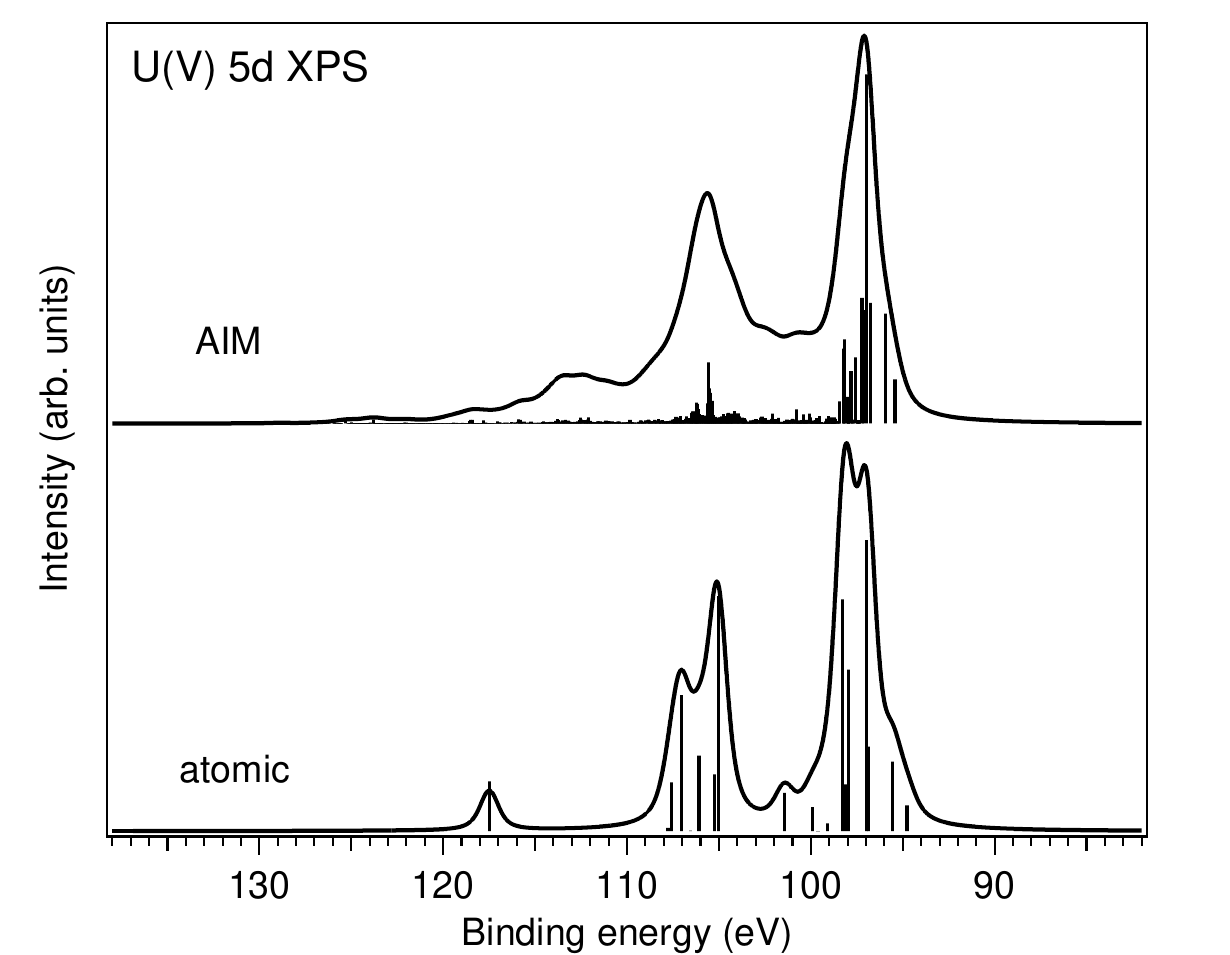}
\caption{Comparison of atomic-multiplet-calculated U(V) $5d$ XPS spectrum with AIM-calculated one for U(V) in the lattice of La-doped UO$_2$. \label{U5_5dxps}}
\end{figure}

To show to what extent the shape of the pre-edge structure in the U(V) $O_{4,5}$ XAS spectra is influenced by the crystal-field interaction, defined by the local environment of the U ion, we calculated these spectra for the octahedral crystal-field strength with $B^{4}_{0}$=2.03 eV and $B^{6}_{0}$=0.17 eV as it was estimated for uranate NaUO$_3$ (Ref. \cite{Butorin_uranates}). The results are represented in Fig.~\ref{U5_Oh_CT}. The relatively strong crystal field leads to a modification of the pre-edge structure as compared to the atomic case but the U $5f$-O $2p$ hybridization causes a further change of this structure.

As a whole, for U(VI) and U(V) oxides, the U $5f$-O $2p$ hybridization and  high covalency degree of the chemical bonding play a decisive role in increasing the $5f$ occupancy and consequently in governing the ground and excited state properties. That is even more pronounced in the calculations of the $5d$ x-ray photoemission spectrum (XPS) of U(V) in the lattice of La-doped UO$_2$. Fig.~\ref{U5_5dxps} shows a comparison of the atomic-multiplet- and AIM-calculated spectra. The reduction factors for the Slater integrals and AIM parameters were the same as in the calculations of the U(V) $O_{4,5}$ XAS spectra (see above). The ground (final) state was described as a mixture of the $5f^{1}$, $5f^{2}\underline{\upsilon}^{1}$ and $5f^{3}\underline{\upsilon}^{2}$ ($5d^{9}5f^{1}$, $5d^{9}5f^{2}\underline{\upsilon}^{1}$ and $5d^{9}5f^{3}\underline{\upsilon}^{2}$) configurations in the AIM calculations. The spectra were broadened by Lorentzian with the full width at half maximum (FWHM) of 0.8 eV and by Gaussian with FWHM=0.6 eV. It is clear from Fig.~\ref{U5_5dxps} that the U $5f$-O $2p$ hybridization and the corresponding charge-transfer lead to major changes in $5d$ XPS. As a result, the AIM-calculated spectrum was also found to be quite similar to the U(V) component derived from the XPS measurements of UO$_{2+x}$ (Ref. \cite{Ilton}).

\section{Conclusions}
The analysis of the actinide $O_{4,5}$ XAS spectra of ThO$_2$ and U(VI) and U(V) oxides, which probe the $5f$ states, indicates that the crystal-field interaction and/or high covalency degree of the chemical bonding are the important factors in defining the electronic structure. An increased $5f$ occupancy due to the U $5f-$O $2p$ charge-transfer influences the ground and excited state properties. While in ThO$_2$ the $5f$ states are mainly affected by the crystal-field interaction with the closest neighbors as can be seen from the corresponding spectrum of ThO$_2$, for U(VI) and U(V) oxides, the main effects in the spectra originate from the strong U $5f$-O $2p$ hybridization.

Another outcome of the $O_{4,5}$ XAS calculations for high-valent light actinides is that Slater integrals $G^k$ for the $5d$-$5f$ interaction need to be reduced to 60\% of their \textit{ab-initio} Hartree-Fock values for actinide ions. This reduction factor is more significant than the conventionally applied one \cite{Ogasawara_5dXAS} and is connected to the increasing degree of the $5f$ delocalization in compounds of high-valent light actinides.
\section{Acknowledgments}
The research was funded by the European Union's "EURATOM" research and innovation program under grant agreement No. 101164053. This work was supported in part by the Director, Office of Science, Office of Basic Energy Sciences, Division of Chemical Sciences, Geosciences, and Biosciences Heavy Elements Chemistry program (D.K.S.) of the U.S. Department of Energy at Lawrence Berkeley National Laboratory under Contract No. DE-AC02-05CH11231. This work used resources of the Advanced Light Source, a U.S. DOE Office of Science User Facility under contract No. DE-AC02-05CH11231.

\bibliography{5dXAS_lightAn}
\end{document}